# Hexadecapolar colloids

Bohdan Senyuk[1], Owen Puls[1], Oleh M. Tovkach[2,3], Stanislav B. Chernyshuk[4] & Ivan I. Smalyukh[1,5,6,*]

[1]*Department of Physics and Soft Materials Research Center, University of Colorado, Boulder, CO 80309, USA*

[2]*Bogolyubov Institute for Theoretical Physics, NAS of Ukraine, Kyiv 03680, Ukraine*

[3]*Department of Mathematics, University of Akron, Akron, OH 44325, USA*

[4]*Institute of Physics, NAS of Ukraine, Kyiv 03650, Ukraine*

[5]*Department of Electrical, Computer, and Energy Engineering, Materials Science and Engineering Program, University of Colorado, Boulder, CO 80309, USA*

[6]*Renewable and Sustainable Energy Institute, National Renewable Energy Laboratory and University of Colorado, Boulder, CO 80309, USA*

*\*email: ivan.smalyukh@colorado.edu*

## Abstract

Outermost occupied electron shells of chemical elements can have symmetries resembling that of monopoles, dipoles, quadrupoles and octupoles corresponding to filled *s*-, *p*-, *d*- and f orbitals. Theoretically, elements with hexadecapolar outer shells could also exist, but none of the known elements have filled *g*-orbitals. On the other hand, the research paradigm of "colloidal atoms" displays complexity of particle behaviour exceeding that of atomic counterparts, which is driven by DNA functionalization, geometric shape and topology and weak external stimuli. Here we describe elastic hexadecapoles formed by polymer microspheres dispersed in a liquid crystal, a nematic fluid of orientationally ordered molecular rods. Because of conically degenerate boundary conditions, the solid microspheres locally perturb the alignment of the nematic host, inducing hexadecapolar distortions that drive anisotropic colloidal interactions. We uncover physical underpinnings of formation of colloidal elastic hexadecapoles and describe the ensuing bonding inaccessible to elastic dipoles, quadrupoles and other nematic colloids studied previously.

## Introduction

Colloids form a platform for scalable fabrication of mesostructured composite materials and provide a framework for testing theoretical descriptions of crystals and glasses, albeit they are also commonly encountered in daily life in forms of milk, paints, coffee, fog and so on[1–5]. Nematic liquid crystal (NLC) colloids[6–17] attract particularly strong interest because they can be tuned by weak external stimuli, such as low-voltage fields and light, similar to the NLC host fluids themselves[12,17]. Colloidal particles locally perturb the uniform ground-state alignment of the anisotropic NLC molecules described by the so-called "director" $\mathbf{n} \equiv -\mathbf{n}$ with non-polar symmetry, producing spatial elastic distortions of the director field $\mathbf{n(r)}$ that resemble electric field configurations around dipolar and quadrupolar charge distributions[7–12]. The type of distortions depends on surface boundary conditions for $\mathbf{n(r)}$, which can be tangential or perpendicular to the particle surfaces and can induce both bulk and surface line and point defects[7–12], topological singularities along which $\mathbf{n(r)}$ and NLC order cannot be defined[1]. Such NLC colloidal particles tend to arrange themselves so that they can share energetically costly $\mathbf{n(r)}$ distortions to minimize free energy, exhibiting highly anisotropic elasticity-mediated interactions that resemble interactions of electrostatic dipoles and quadrupoles[6–12,15,16]. This electrostatics analogy[7] provides a framework for understanding, predicting and engineering the ensuing colloidal self-assembly. However, for many years of very active research efforts[6–12,15,16], only colloidal elastic dipoles and quadrupoles have been found in experiments and theories alike, which limits diversity of the accessible colloidal bonds and self-assembled structures[6–17].

In this work, we describe colloidal elastic hexadecapoles that spontaneously form around solid microspheres immersed in a uniformly aligned NLC. The unusual symmetry of elastic distortions arises from conically degenerate[18] boundary conditions for $\mathbf{n(r)}$ at the colloidal surfaces, which induce surface point and line defects at the same time. Using a combination of holographic optical tweezers (HOT)[19], nonlinear optical imaging[20] and polarizing optical microscopy (POM)[1,6], we probe the $\mathbf{n(r)}$-distortions and quantify colloidal pair interactions by measuring distance and angular dependencies of elastic potentials, demonstrating relations between the director structure and medium-mediated inter-particle forces. Finally, we explain our findings using a model based on elastic multipole expansion and discuss how the experimental framework we have developed may enable colloidal self-assembly into novel forms of tunable pre-engineered matter without known atomic analogues.

## Results

**Hexadecapolar elastic multipole.** When dispersed in a uniformly aligned NLC fluid host, polystyrene microspheres (PSMs) of a radius $r_0$ locally distort $\mathbf{n}(\mathbf{r})$, which is manifested by eight bright lobes around the particle perimeter seen in POM between the crossed polarizer and analyser (Fig. 1a). These bright lobes are separated by eight dark regions within which $\mathbf{n}(\mathbf{r})$ at the particle's perimeter is parallel to polarizer or analyser. Using a phase retardation plate and interference of polarized light propagating through the particle-distorted structure, we reveal that $\mathbf{n}(\mathbf{r})$ tilting away from the far-field director $\mathbf{n}_0$ switches between clockwise and counterclockwise directions (corresponding to the blue and yellow colours in the micrograph) eight times as one circumnavigates the sphere (Fig. 1b). Bright-field micrographs obtained at different depth of focus reveal presence of weakly scattering surface point defects (boojums) at the particle poles along $\mathbf{n}_0$ as well as a circular loop of a defect line (often called "Saturn ring")[9,10] at the particle's equator (Fig. 1c,e). Based on POM and three-dimensional nonlinear optical imaging (Fig. 1a–c,e and Supplementary Figs 1–3), we uncover the structure of $\mathbf{n}(\mathbf{r})$ distortions schematically shown in Fig. 1d. This structure is consistent with conically degenerate surface boundary conditions for $\mathbf{n}(\mathbf{r})$ with respect to PSM's local surface normals $\mathbf{s}$, which were previously demonstrated for NLCs at flat polystyrene-coated surfaces[18]. The director's easy axis orientation lies on a cone of equilibrium polar angle $\psi_e$ (Fig. 1f). To minimize the free-energy cost of bulk elastic distortions, interaction of conically degenerate surface boundary conditions on the microsphere with the uniform $\mathbf{n}_0$ lifts this conical degeneracy and yields an axially symmetric $\mathbf{n}(\mathbf{r})$ depicted in Fig. 1d. Extending this analysis to three dimensions, the projection $n_x$ of $\mathbf{n}(\mathbf{r})$ onto the $x$-axis orthogonal to $\mathbf{n}_0$ can be visualized around PSM using colours that highlight positive, near zero and negative $n_x$ (Fig. 1g). The black points at the poles and ring at the equator of the sphere are regions of discontinuity of $\mathbf{n}(\mathbf{r})$ at the NLC-PSM interface and correspond to the boojum and Saturn ring topological defects, respectively. Away from the particle surface and these topological singularities, the experimentally reconstructed $\mathbf{n}(\mathbf{r})$ is continuous (Fig. 1d) and consistent with the theoretical model for a hexadecapolar director distortions presented in a similar way in Fig. 1h.

**Elastic interactions between hexadecapolar colloids.** Elasticity mediated interactions between PSMs (Fig. 2) differ from all NLC colloids studied so far. To get insights into the strength and direction-dependence of these interactions, we utilize HOT to optically trap one "stationary" PSM and then release another particle at different centre-to-centre vector **R** orientations with respect to $\mathbf{n}_0$ as well as at different separation distances. Using video-microscopy, we track the ensuing particle motions, which result from a combination of Brownian jiggling and elasticity mediated interactions. The submicron waist and relatively low power (~ 10 mW) of a focused trapping beam allow us to avoid the influence of the trapping on the measurements[21]. Furthermore, the laser tweezers are used only to bring the particles to the desired initial conditions and are turned off within the time when the pair-interaction is probed with video microscopy, allowing us to avoid possible artefacts associated with the complex effects of the laser trapping light at small interparticle distances. The colour-coded time-coordinate trajectories of particles released from optical traps at different **R** are shown in Fig. 2a. Unlike in the case of dipolar and quadrupolar NLC colloids[7,8,15,16], elastic forces are relatively short-ranged and exceed the strength of thermal fluctuations only at distances of four-to-five particle radii $r_0$. However, the angular dependence of these forces is very rich, with eight angular segments of inter-particle attractions separated by eight angular segments of repulsions, with the intermediate angular sectors within which particles move sideways as the inter-particle elastic forces are orthogonal or at large angles to **R** (Fig. 2a). These angular sectors of attraction and repulsion correlate with the bright and dark regions of POM micrographs (Fig. 1a) as well as with the predictions of our model based on elastic multipole expansion (Fig. 2).

To quantify elastic interactions, we first probe the centre-to-centre separation $R = |\mathbf{R}|$ versus time $t$ for particles released at different angles between **R** and $\mathbf{n}_0$ within the angular sectors of attraction (Fig. 3a) and then calculate particle velocities $\dot{\mathbf{R}} = d\mathbf{R}/dt$. Because the system is highly over-damped and the inertia effects can be neglected[22], the experimentally measured $\mathbf{R}(t)$ and the simplified equation of motion $0 \approx -\xi\dot{\mathbf{R}} + \mathbf{F}_{\text{int}}$ yield the pair interaction potential $U_{\text{int}}$ (inset of Fig. 3a), where $\xi$ is a drag coefficient measured separately by characterizing Brownian motion of PSMs (Methods and Supplementary Fig. 4) and $\mathbf{F}_{\text{int}} = -\nabla U_{\text{int}}$ is the elastic interaction force. The attractively interacting particles stop short of physically touching each other, instead forming stable dimer assemblies with typical $R \approx (2.05\text{-}2.2)r_0$ and stable **R** orientations with respect to $\mathbf{n}_0$ within one of the two angular sectors of assembly in each quadrant dependent on $\psi_6$: $\theta_1 \approx 22°\text{-}26°$ or $\theta_2 \approx 64°\text{-}75°$ (Fig. 3b–f). Multi-particle assemblies with different combinations of angles $\theta_1$ and $\theta_2$ are also observed (Fig. 3d). The inter-particle binding energies are measured to be in the range of hundreds of $k_BT$ (inset of Fig. 3a), making them robust with respect to thermal fluctuations. Although $U_{int}$ versus $\theta$ has eight minima, only four of them can be occupied simultaneously in one plane by particles of the same size because of the "excluded volume" effects, yielding two-dimensional colloidal crystals with rhombic elementary cells (Supplementary Fig. 5 and Supplementary Note 1). Following similar considerations, a large number of low-symmetry three-dimensional colloidal structures can be envisaged too. Since the elastic interactions potential is hundreds of $k_BT$ and the particle assemblies can be entrapped in metastable states, the assembly of two- and three-dimensional colloidal lattices from micrometer-sized particles requires the use of optical tweezers for guiding colloidal particles. Alternatively, the elastic interaction potentials between colloidal particles of smaller size or with weaker surface anchoring can be brought to the order of $10\ k_BT$ and lower, so that the crystal self-assembly can occur without the assistance of optical tweezers, which will be explored elsewhere.

To model experiments, we exploit the electrostatic analogy of the far-field director distortions because of a colloidal sphere that can be represented as elastic multipoles[23], albeit our colloids dramatically differ from elastic dipoles and quadrupoles studied so far[7–12,23–26]. Far from the colloidal sphere, the director deviations $n_\mu$ ($\mu = x; y$) from $\mathbf{n}_0 = (0, 0, 1)$ are small. Assuming $\mathbf{n}(\mathbf{r}) \approx (n_x, n_y, 1)$, the NLC elastic free-energy reads[7,8,23]

$$F_{har} = \frac{K}{2} \sum_{\mu=x,y} \int d\mathbf{r} \nabla n_\mu \cdot \nabla n_\mu ,$$ (1)

where $K$ is an average Frank elastic constant[1]. Euler-Lagrange equations arising from the functional (1) are of Laplace type, $\Delta n_\mu = 0$, with solutions expanded into multipoles

$$n_\mu(\mathbf{r}) = \sum_{l=1}^{N} a_l (-1)^l \partial_\mu \partial_z^{l-1} \frac{1}{r} ,$$ (2)

where $a_l = b_l r_0^{l+1}$ is the elastic multipole moment of the $l$th order ($2^l$-pole)[22–26] and one can find coefficients $b_l$ from exact solutions for $\mathbf{n}(\mathbf{r})$ or from relevant experiments (Figs 2a and 3a). Odd moments vanish because $\mathbf{n}(\mathbf{r})$ is symmetric about the particle centre, similar to analogous electrostatic charge distributions that have no dipole or octupole electric moments[22–26]. The multipole expansion of the induced $\mathbf{n}(\mathbf{r})$ shown in Fig. 1d is characterized by coefficients $b_2$, $b_4$ and $b_6$, which also determine the colloidal pair-interaction energy[24]

$$U_{\text{int}} = 4\pi K \sum_{l,l'=2,4,6} a_l a'_{l'} (-1)^{l'} \frac{(l+l')!}{R^{l+l'+1}} P_{l+l'}(cos\,\theta),\qquad(3)$$

where $P_{l+l'}(cos\,\theta)$ are the Legendre polynomials. For colloidal quadrupoles, $b_2$ dominates and $b_4$ and $b_6$ can play a role only at small[25] $R$. For our particles (Fig. 1), the induced $\mathbf{n}(\mathbf{r})$ can be qualitatively understood as a superposition of configurations of two separate quadrupoles, one with the Saturn-ring and one with boojums[9–11,16,22–26], having opposite signs of quadrupole moments (compare Fig. 4a,d,g,j and 4b,e,h,k). Therefore, the net quadrupole moment is small and the high-order multipoles manifest themselves in a wide range of $R$. Fitting experimental $R(t)$ within different angular sectors of pair interaction with the corresponding theoretical predictions yields a unique set of parameters $b_2$, $b_4$ and $b_6$ (Fig. 3a). The quadrupole moment $a_2 = -0.017 r_0^3$ is about two orders of magnitude smaller than that of elastic quadrupoles[16,24], consistent with the hexadecapolar symmetry of $\mathbf{n}(\mathbf{r})$ that is found playing a dominant role, and even the higher order term (64-pole) plays a detectable role at relatively small $R$ (Supplementary Fig. 6 and Supplementary Note 2). The rich angular dependence of elastic pair-interactions predicted by equation 3 for coefficients $b_2$, $b_4$ and $b_6$ obtained from fitting is consistent with our experimental characterization of the hexadecapolar nature of PSM colloids in the NLC host (Fig. 2).

## Discussion

It is instructive to compare hexadecapolar NLC colloids to other known elastic multipoles and their electrostatic analogues. Colloidal spheres with strong tangential anchoring at their surface form $\mathbf{n}(\mathbf{r})$-distortions of quadrupolar configuration[16,24,27,28] and are commonly called elastic quadrupoles (Fig. 4a,d,g,j). They have two surface point defects at the two poles of the spherical particle, which called "boojums". Spherical particles with homeotropic surface anchoring can induce two different elastic multipoles, a dipole[23,28] (Fig. 4c,f,l) and a quadrupole[16,23,27–32] (Fig.

4b,e,h,k). The elastic colloidal dipole has a bulk point defect called "hedgehog" near one of the poles and lacks mirror symmetry with respect to the plane going through its equator orthogonally to $\mathbf{n}_0$ (Fig. 4f,l). Such a colloidal dipole has a small octupole moment as well[33]. The elastic quadrupole around the particle with a strong homeotropic anchoring has a closed disclination loop around the equator called "Saturn ring" and, in all four quadrants, has an opposite tilt of the deformed director with respect to $\mathbf{n}_0$ as compared with the tangential elastic quadrupole[16] (compare Fig. 4g,j,h,k). In the case when anchoring is weak, an elastic multipole with weak distortions of quadrupolar configuration can be formed somewhat similar to the elastic quadrupole with "Saturn ring", but its disclination loop is "virtual" (within particle's volume) as the director is allowed to deviate away from the easy axis orientation[31,32]. The maps of $x$-component of the director, $n_x$, calculated using methods described in refs 23,24 and plotted on spherical surfaces encompassing the particles and the defects, clearly illustrate the multipolar nature of elastic distortions similar to that in the electrostatics analogues[34,35]. A direct comparison of the symmetry of the director distortions around elastic dipoles and quadrupoles with that of hexadecapolar PSMs in NLCs helps to highlight the very different nature of these high-order NLC colloidal multipoles studied in our work (compare Figs 4j–l and 1h). On the other hand, the hexadecapolar NLC colloids further expand the analogy between electrostatic and elastic colloidal multipoles. In fact, the colour presentations of elastic distortions induced by colloidal dipoles (Fig. 4l), two different quadrupoles (Fig. 4j,k) and our hexadecapoles (Fig. 1g,h) resemble very closely the corresponding dipolar, quadrupolar and hexadecapolar electrostatic charge distributions on a sphere described by a spherical harmonic function[35] $\sigma_l^m(\Theta, \phi) = N\cos(m\phi)P_l^m(\cos\Theta)$ with $(l, m) = (1, 1)$, $(2, \pm1)$ and $(4, 1)$, respectively, where $N$ is a normalization constant, $\Theta$ is a polar and $\phi$ is an azimuthal angles, $P_l^m(\cos\Theta)$ is the associated Legendre polynomial; the constant $l$ determines the order of a multipole (that is, $2^l$-th pole) and $-l \leq m \leq l$. This close analogy between electrostatic and elastic colloidal multipoles may help devising approaches for self-assembly of colloidal mesostructured composite materials.

To conclude, we describe hexadecapolar NLC colloids with unusual field configurations, highly anisotropic elastic interactions, and versatile forms of self-assembly. Our findings pose a challenge of realizing pure NLC colloidal octupoles, lower order multipoles that have not been observed so far, and provide the means of probing the role of hexadecapolar moments in inter-molecular interactions using colloidal model systems. Self-assembly of hexadecapolar NLC colloids is expected to yield a diverse family of two- and three-dimensional low-symmetry crystal lattices. Beyond the rich experimental platform for fundamental studies, our colloids have potential uses in designing and realizing reconfigurable "soft" photonic crystals[36] and other NLC-based mesostructured composites[37] with properties that can be pre-engineered through controlling tilted boundary conditions on nanoparticle surfaces[38–40], particle shape, composition and topology[2,6,8,41], as well as the NLC host properties, particle active behaviour[42] and the use of external stimuli[12,17].

## Methods

**Preparation of nematic colloidal dispersions.** Colloidal PSM particles of radii $r_0 \approx 2.05$, 2.65 and 4 μm and with conically degenerate surface boundary conditions were prepared using one of two different methods as follows[43,44]. In the first method, asymmetric colloidal polystyrene dimers[43,45] were synthesized using a modified seeded polymerization technique[44–46] (Supplementary Fig. 1a). Resulting dimers in ethanol were sonicated for tens of minutes using an ultrasonic bath 8891 (Cole-Parmer) to break apart large and small spherical lobes forming the asymmetric colloidal dimer (Supplementary Fig. 1b,c). Colloidal spheres originating from these dimers were then re-dispersed in pentyl-cyano biphenyl (5CB) obtained from Frinton Laboratories, Inc. The larger (diameter $D = 2r_0 \approx 4.1$ μm) colloidal spheres, which were used as PSMs in our experiments, intrinsically provide conic degenerate surface boundary conditions for the NLC director. In the second preparation method, we used polystyrene divinylbenzene spherical particles PC06N (Bang Laboratories, Inc.) of $D \approx 8$ μm and DC-05 (Thermo Scientific) of $D \approx 5.3$ μm dispersed in 5CB. As dispersed, these particles provide tangential alignment for 5CB molecules and the director field $\mathbf{n(r)}$. However, after keeping these colloidal dispersions in 5CB at the elevated temperature of $\approx 100°C$ for about 12 h, the alignment at the surface of these polymer particles changed to a conical degenerate (Supplementary Fig. 1d–f), similar to that reported in ref. 18 for flat surfaces coated by polystyrene. The tilt angle of the easy axis $\psi_e$ was varying approximately within 20°-75°, depending on details of preparation. The comparison of the surface free-energy density $F_s = W_c(\cos^2\psi - \cos^2\psi_e)^2/2$ costs (ref. 18) for deviating $\mathbf{n(r)}$ to a polar angle $\psi$ away from the easy axis at $\psi_e$ relative to the bulk elastic energy density is characterized by the so-called conical anchoring extrapolation length $K/W_c \sim$ (50-500) nm $<< r_0$, where[18] $W_c \sim 10^{-5}$ J m$^{-2}$. Since $K/W_c << r_0$, one can assume that the tilted conically degenerate boundary conditions are rigid with respect to changing $\psi$. To prepare samples with NLC dispersions in a uniformly aligned host, dry PSMs were added to 5CB. The dilute NLC colloidal dispersions were then filled into the cells comprised of two glass plates separated by glass spacers defining the cell gap in the range 15-30 μm, which were sealed at all edges using fast-setting epoxy glue after the filling process. The inner surfaces of cell substrates were treated with polyimide PI2555 (obtained from HD Microsystems) for in-plane homogeneous alignment of the far-field director $\mathbf{n}_0$ defined by unidirectional rubbing.

**Imaging techniques and data acquisition.** POM (Fig. 1 and Supplementary Figs 1,2), nonlinear optical imaging (Supplementary Fig. 3) and laser manipulations of PSM particles were performed using a single integrated setup capable of simultaneous conventional optical and nonlinear three-photon excitation fluorescence polarizing microscopy (3PEF-PM)[20] and HOT optical control operating at 1,064 nm (ref. 19). The setup was built around an inverted microscope IX81 (Olympus). For 3PEF-PM imaging[20], we have employed a tunable (680-1,080 nm) Ti-Sapphire oscillator (Chameleon Ultra II, Coherent) emitting 140 fs pulses at a repetition rate of 80 MHz. The laser wavelength was tuned to 870 nm for the three-photon excitation of 5CB molecules and the 3PEF-PM signals were collected in epi-detection mode with a photomultiplier tube (H5784-20, Hamamatsu). An Olympus ×100 oil-immersion objective with a high numerical aperture of 1.4 was used for both optical imaging and laser trapping. This experimental setup is described in details elsewhere[19,20]. Motion of colloidal particles was recorded with a charge-coupled camera (Flea, PointGrey) at a rate of 15 fps and their lateral positions versus time were determined from captured sequences of images using motion tracking plugins of ImageJ software (obtained from National Institutes of Health) with the accuracy[8] of 7-10 nm.

**Measuring anisotropic diffusion of elastic NLC hexadecapoles.** PSMs with conic anchoring distort the homogeneous $\mathbf{n}(\mathbf{r})$ and form elastic hexadecapoles in the aligned NLC (Fig. 1a–e and Supplementary Figs 1b,d–f, 2,3b). These hexadecapoles stay suspended in the bulk of NLC (Supplementary Fig. 3b), which is facilitated by elastic repulsion of PSMs from the confining substrates and their Brownian motion because of thermal fluctuations. Using bright field microscopy and video tracking, one can determine a position of the particle within each frame and then analyse its translational displacements at regular time steps corresponding to the video frame rates. Following the well-established methodology[29] and using the histograms of displacements shown in Supplementary Fig. 4, it is possible to experimentally determine two independent diffusion coefficients $D_{\parallel} = 1.63 \times 10^{-3}$ $\mu m^2$ $s^{-1}$ and $D_{\perp} = 1.05 \times 10^{-3}$ $\mu m^2$ $s^{-1}$, which characterize, respectively, the diffusion of elastic hexadecapoles along and perpendicular to the far-field director $\mathbf{n}_0$. As $D_{\parallel} > D_{\perp}$, it is easier for the NLC hexadecapoles to move along $\mathbf{n}_0$ than in other directions. The ratio $D_{\parallel} > D_{\perp} = 1.55$ is not much different from values determined for other elastic multipoles studied previously[29,30] (Fig. 4). These coefficients are also used to determine friction coefficients from the Stokes-Einstein relation, which are then subsequently used in the calculation of anisotropic interaction forces[27] between hexadecapoles.

**Acknowledgements**

We thank T. Lubensky, S. Wang and N. Wu for discussions and acknowledge support of the National Science Foundation Grant DMR-1410735.

**Author contributions**

B.S. and O.P. performed the experimental work. B.S., O.M.T., S.B.C. and I.I.S. analysed the data. I.I.S. and B.S. wrote the manuscript with an input from all co-authors. I.I.S. conceived and directed the project.

**Figures**

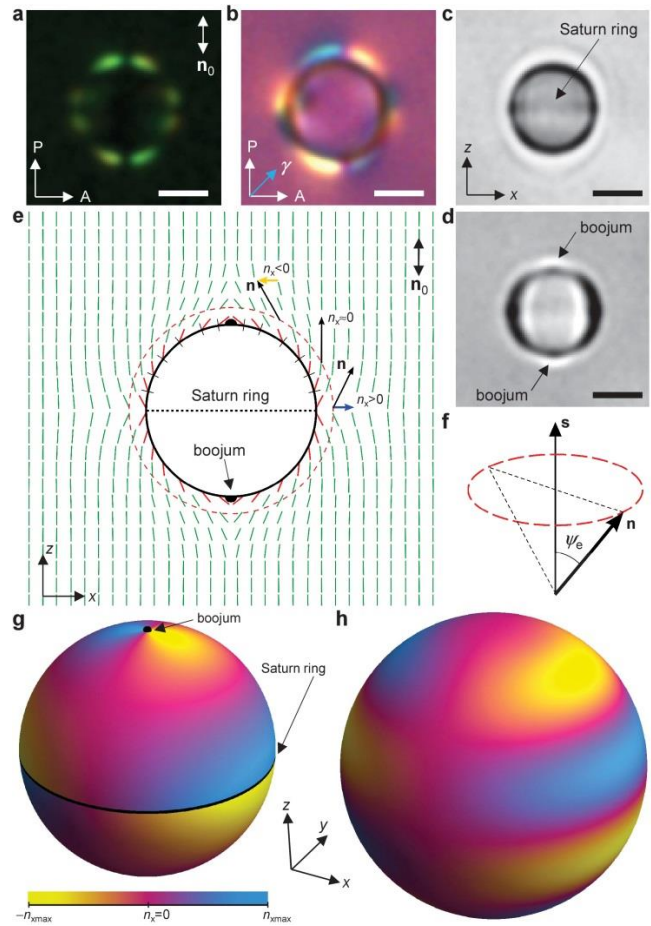

**Figure 1 | Elastic hexadecapole induced by a colloidal PSM. (a–d)** Optical micrographs obtained with **(a,b)** POM and **(c,e)** bright field microscopy. *P*, *A* and *γ* mark the crossed polarizer, analyser and a slow axis of a 530 nm retardation plate (aligned at 45° to *P* and *A*), respectively. **(d)** Schematic diagram of induced $\mathbf{n(r)}$ (green rods) satisfying the tilted boundary conditions at the PSM surface (red rods), with the "easy axis" at a constant angle $\psi_e$ to a local normal $\mathbf{s}$ to the surface (black rods). **(f)** Schematic of conic degenerate surface boundary conditions. **(g,h)** Three-dimensional visualization of the *x*-component $n_x$ of $\mathbf{n(r)}$ **(g)** at the surface of PSM for $\psi_e = 45°$ and **(h)** at a spherical surface of radius $1.2r_0$ shown using a dashed red circle in **d**. Blue, yellow and magenta colours correspond, respectively, to a positive, near-zero and negative $n_x$ according to the colour scheme shown in **g**. Dashed equatorial line in **d** and a black solid line in **g** depict the "Saturn ring" surface defect loop at the particle's equator visible in **c**. Black hemispheres in **d** and **g** show surface point defects boojums at the poles of the particle visible in **e**. Scale bars, 2 μm.

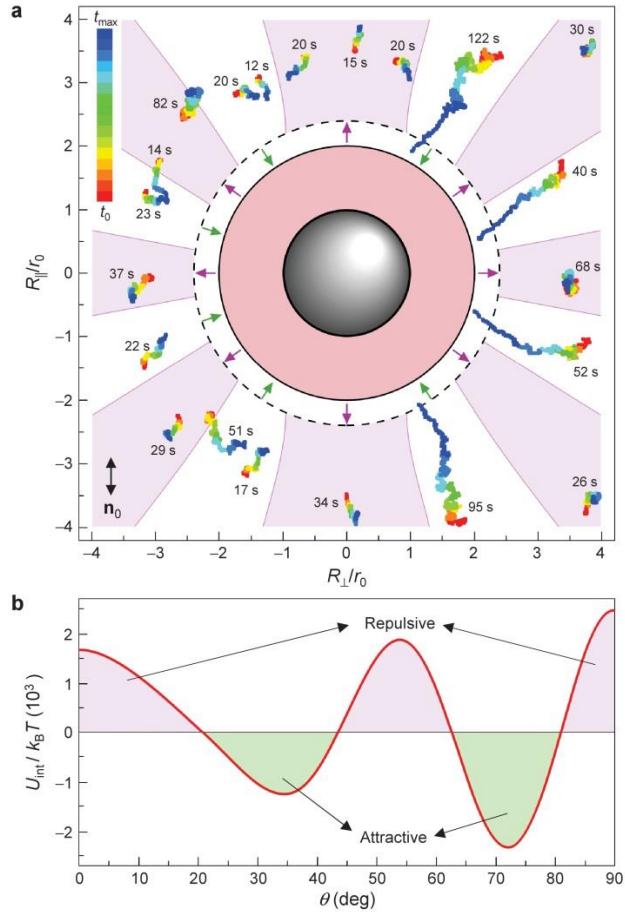

**Figure 2 | Elastic interactions of colloidal hexadecapoles.** (**a**) Angular dependence of interactions is probed by tracking motion of a particle released from the optical trap and moving with respect to the "stationary" trapped particle in the centre depending on the orientation of **R** with respect to $\mathbf{n}_0$. The elapsed time is coded according to the colour scale (inset) and the maximum elapsed times $t_{max} - t_0$ are marked next to the corresponding trajectories. $R_{\parallel}$ and $R_{\perp}$ denote the centre-to-centre distance **R** components along and perpendicular to $\mathbf{n}_0$, respectively. The PSM of radius $r_0$ is surrounded by the spherical volume of radius $2r_0$ that is excluded for centres of other PSMs. The nonlinear zone is shown by a dashed circle at $R = 2.4r_0$. (**b**) Pair-potential $U_{int}$ versus y for two particles at $R = 2.4r_0$. Angular zones of repulsion and attraction are highlighted using magenta and green arrows and colouring, respectively.

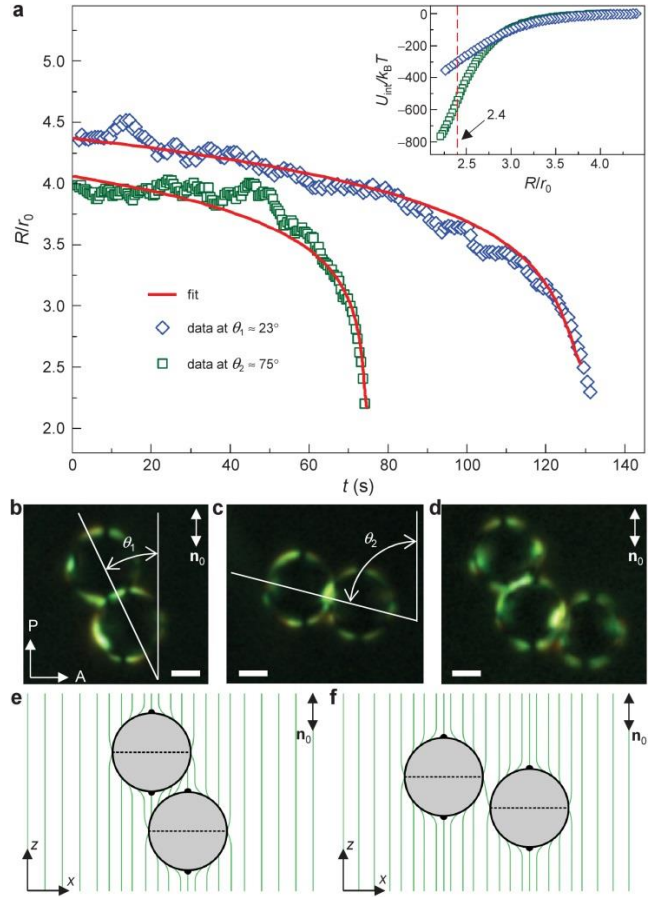

**Figure 3 | Self-assembly of hexadecapolar colloids.** (**a**) Separation $R$ versus time $t$ for attractive interactions at different $\theta$ (measured upon the dimer formation). The solid red line shows a least-squares fit with $R(t)$ obtained from the simplified equation of motion with ($b_2$, $b_4$, $b_6$) = (-0.017, -0.092, 0.003); it is impossible to reproduce such a dependence $R(t)$ with the parameter $b_2$ only since the quadrupole-quadrupole interaction is repulsive along $\theta \approx 23°$. Inset shows the corresponding $U_{int}$ versus $R$. (**b**) and (**c**) POM micrographs of self-assembled colloidal dimers of hexadecapoles with **R** aligned at $\theta_1$ and $\theta_2$ to $\mathbf{n}_0$, respectively. (**d**) POM micrograph of a kinked colloidal chain of hexadecapoles. (**e,f**) Schematic diagrams of $\mathbf{n}(\mathbf{r})$ (green lines) for colloidal dimers shown in **c,d**, respectively. Dashed lines and black hemispheres in **c** and **d** depict surface defect loops and surface point defects at the poles of PSMs, respectively. Scale bars, 2 μm.

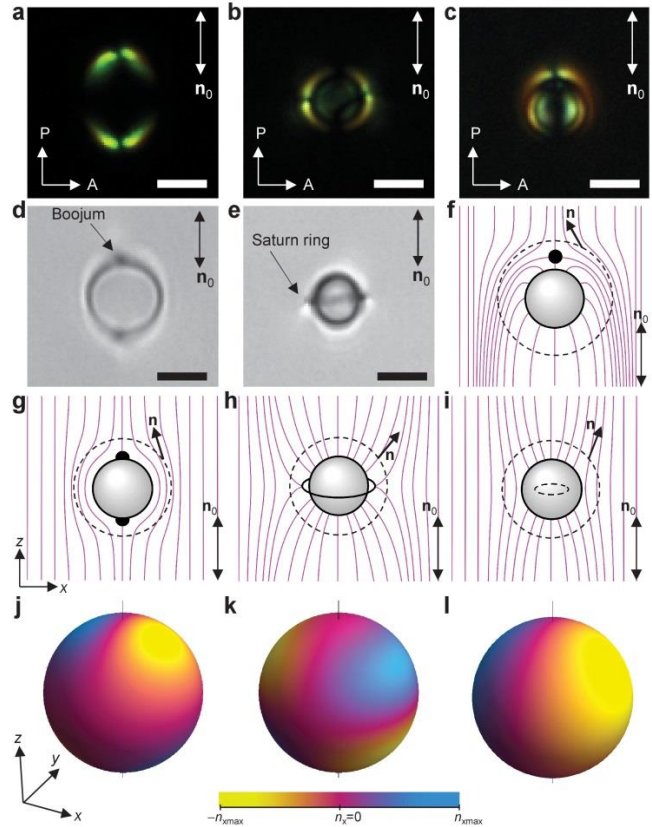

**Figure 4 | Elastic multipoles induced by spherical particles with different anchoring.** (**a**,**b**) Polarizing and (**d**,**e**) bright field micrographs and (**g**,**h**) schematic diagrams of a director field **n(r)** (magenta lines) around elastic quadrupoles, respectively, formed by spheres with a strong (**a**,**d**,**g**) tangential and (**b**,**e**,**h**) homeotropic anchoring. Black filled semicircles at the poles of the particle in **g** show surface point defects dubbed "boojums." A line at the particle equator in **h** shows a closed surface disclination loop dubbed "Saturn ring." (**c**) Polarizing micrograph and (**f**) a schematic diagram of **n(r)** around an elastic dipole formed by a sphere with a strong homeotropic surface anchoring. A black point in the front of the particle shown schematically in **f** depicts a bulk point defect "hedgehog." (**i**) Schematic diagram of a quadrupolar **n(r)** around a sphere with weak surface anchoring. A dashed loop inside the particle shows a virtual disclination ring. (**j**–**l**) Colour-coded visualizations of the $n_x$ component of **n(r)**, which is caused by the tilt of **n(r)** away from **n_0**, at the spherical surface enveloping the particles shown, respectively, in **f**,**g**,**h**. The spherical surfaces of colour-coded visualization of director distortions have radius of $(1.2-1.4)r_0$, as shown by dashed circles in **f**–**h**. Blue, yellow and magenta colours correspond, respectively, to a positive, zero and negative $n_x$ according to a colour scale bar provided in **k**. Scale bars, 4 μm.

## Supplementary Figures

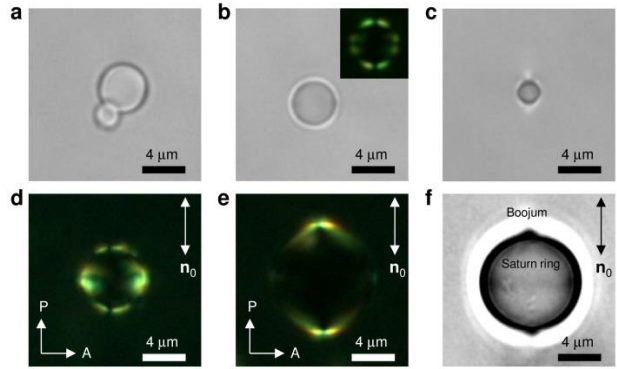

**Supplementary Figure 1 | Preparation of colloidal PSMs with conically degenerate anchoring. (a-c)** Optical bright field micrographs showing two initially connected spherical lobes of (**a**) an asymmetric colloidal dimer, which yield (**b**, **c**) two spherical particles after breaking apart, including the larger one (**b**), which spontaneously exhibits conically degenerate surface anchoring; particles in **a-c** are imaged on a glass substrate. Inset of **b** shows a POM micrograph of an elastic hexadecapole induced by the larger particle when dispersed in 5CB. (**d**, **e**) Polarizing and (**f**) bright field optical micrographs of the polystyrene divinylbenzene particles of different diameter with conic anchoring obtained after the thermal treatment described in Methods.

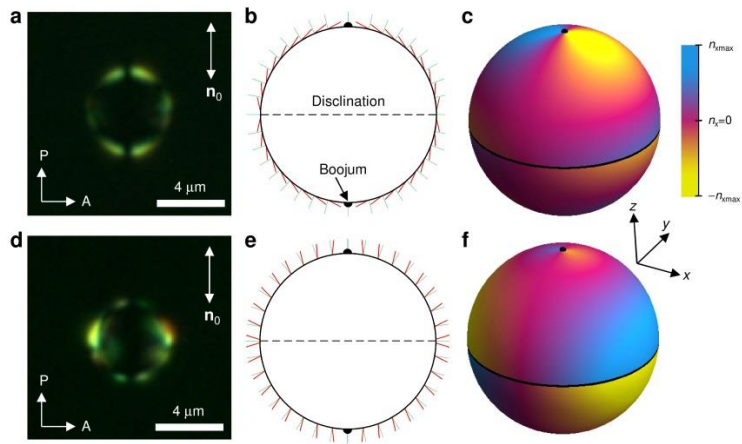

**Supplementary Figure 2 | Director field around PSMs with conically degenerate surface anchoring.** (**a**, **d**) POM micrographs of elastic hexadecapoles formed around spheres with conic anchoring of a different tilt. (**b**, **e**) Schematic diagrams of the local easy axis for the director (red rods) orientation at the surface of the sphere in the case of (**b**) a relatively large and (**e**) relatively small tilt away from the local surface normal (green rods). (**c**, **f**) Color-coded maps of the director distortions showing $n_x$ at the surface of particles shown, respectively, in **a**, **b** and **d**, **e**.

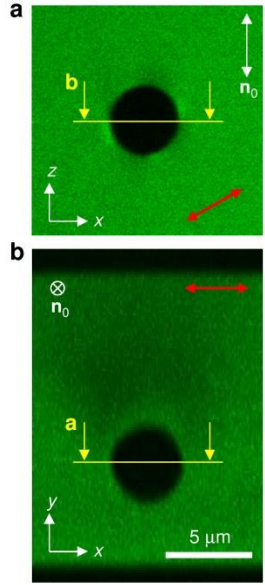

**Supplementary Figure 3 | 3PEF-PM images of an elastic hexadecapole in 5CB.** The images were obtained in the planes marked on the corresponding cross-sections in **a** and **b**. Red double arrow shows a direction of polarization of the excitation beam.

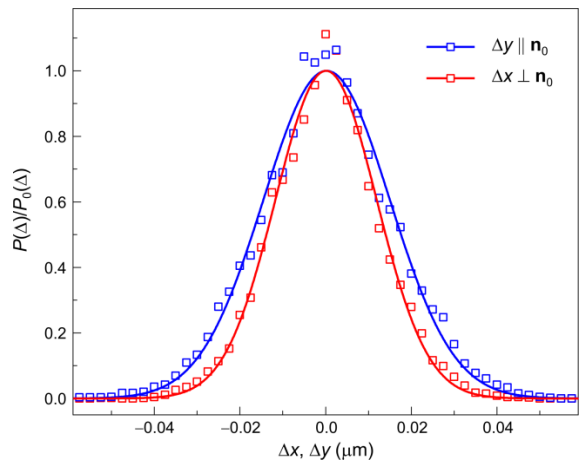

**Supplementary Figure 4 | Normalized histograms of hexadecapole's displacements**. Displacements along ($\Delta y$) and perpendicular ($\Delta x$) to the far-field director $\mathbf{n}_0$ were characterized at the frame rate of 15 fps. The open square symbols show experimental data and the solid lines are the corresponding Gaussian fits.

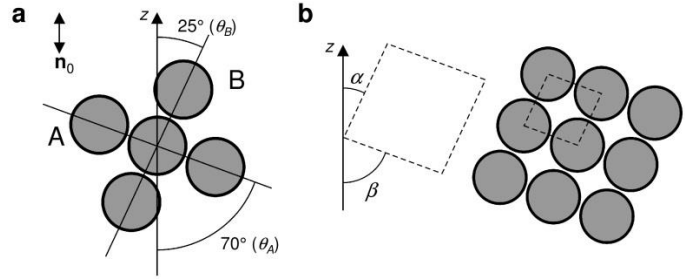

**Supplementary Figure 5 | Two-dimensional assembly of hexadecapolar colloids.** (**a**) Because of the "steric" effects combined with anisotropic interactions, every particle can have no more than four neighbors. (**b**) Planar colloidal lattice with a rhombic elementary cell shown by the dashed lines. Its free energy depends on the orientation which can be unambiguously defined by the angles $\alpha$ and $\beta$, with the anisotropic interactions expected to yield minima at $\alpha \approx \theta_1$ and $\beta \approx \theta_2$.

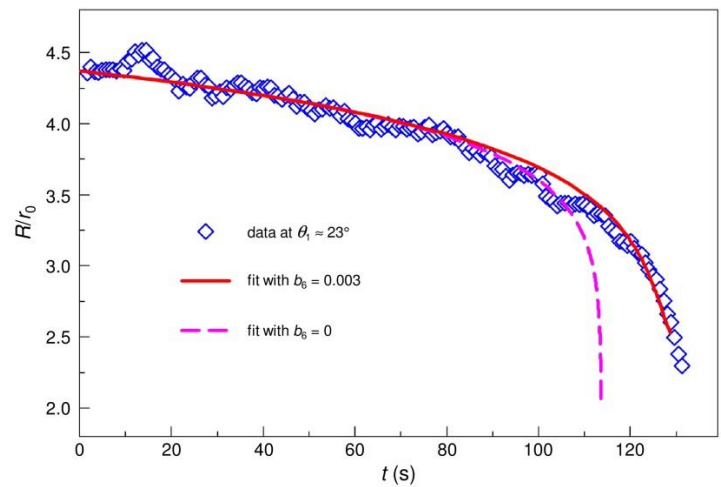

**Supplementary Figure 6 | Role of the higher order terms in fitting the separation versus time data.** Center-to-center particle separation $R$ along the direction at $\theta_1 \approx 23°$ (average angle on contact) with respect to $n_0$. The dashed magenta line shows a least-squares fit with $R(t)$ obtained from the simplified equation of motion with only two adjustable parameters $(b_2, b_4) = (-0.026, -0.071)$. The solid red line corresponds to the solution with three adjustable parameters $(b_2, b_4, b_6) = (-0.017, -0.092, 0.003)$.

# Supplementary Notes

## Supplementary Note 1 - Analysis of colloidal lattices

To explore the structural possibilities for colloidal self-organization of elastic hexadecapoles, let us first consider a two-dimensional system. The pairwise interactions between the hexadecapolar NLC colloids are highly anisotropic. At a fixed inter-particle distance, the interaction energy $U_{int}$ as a function of the angle $\theta$ has eight minima (Fig. 2). However, since the particles are not pointwise objects and exclude volume around them that becomes inaccessible to other particles, only four of these minima can be occupied simultaneously (Supplementary Fig. 5). As a result, every colloidal particle has four nearest neighbors with the approximate arrangements with respect to $\mathbf{n}_0$ depicted in the Supplementary Fig. 5a. Arrangements that are mirror-symmetric with respect to $\mathbf{n}_0$ as well as the ones with inversion symmetry are also possible because of the non-polar uniaxial symmetry of the NLC host fluid. Upon formation of such a five-particle assembly, similar geometric constraints on attractive interactions of the peripheral particles within the assembly lead to additions of new particles at well-defined peripheral sites and eventually prompt formation of a regular two-dimensional lattice with a rhombic elementary cell (Supplementary Fig. 5b). Minimization of the cell's energy over the angles $\alpha$ and $\beta$ defined in the Supplementary Fig. 5b yields their values, $\alpha \approx \theta_1$ and $\beta \approx \theta_2$, that correspond to the minima of the pairwise interaction determined in the main text. Hence, the ensuing colloidal crystal structure is expected to be driven mainly by interactions between the nearest neighbors. Extension of this analysis based on pair-interactions to three dimensions points into the possibility of interesting three-dimensional colloidal assemblies. For example, the planar five-particle assembly shown in Supplementary Fig. 5a can be readily transformed into a tetrahedron via rotation of the particles A and B around the $z$-axis by approximately 90°. Under such a transformation, $\theta_A$ and $\theta_B$ do not change and the energy of the colloidal unit cell remains close to its minimum. Hypothetically, the tetrahedrons can be subsequently assembled into a three-dimensional crystal lattice. Given the angular dependence of $U_{int}$ and the shape of two-dimensional elementary cell, one can also expect existence of three-dimensional colloidal crystals with low-symmetry (e.g. triclinic) lattices. Our highly over-simplified analysis here is of course based on pair interactions, but many-body and kinetic effects may become important in defining three-dimensional assemblies and will be of great interest to explore in future studies.

## Supplementary Note 2 - Interaction potential of elastic hexadecapoles

It is well established that the far-field nematic director distortions induced by colloidal particles can be represented by means of elastic multipoles[1,2]. One can find unknown multipole moments either from long-range asymptotic of exact solutions for $\mathbf{n(r)}$ or from relevant experiments. Rigorously speaking, the multipole expansion given by Eq. (2) is an infinite series. In order to determine where it should be truncated, one can estimate the leading anharmonic correction $F_{anh}$ to the free energy given by Eq. (1) associated with $\nabla n_z$. Since the $F_{har}$ is derived under the assumption of $n_z \approx 1$, one finds that $F_{anh} = \frac{K}{2} \int d\mathbf{r} (\nabla n_z)^2 \approx \frac{K}{8} \int d\mathbf{r} (\nabla n_\perp^2)^2$, where $n_\perp^2 = n_x^2 + n_y^2$. This modifies the Euler-Lagrange equations to become

$$\Delta n_\mu + \frac{1}{2} n_\mu \Delta n_\perp^2 = 0.\tag{4}$$

It follows from Eq. (4) that if the leading term of the multipole expansion falls off as $r^{-n}$ then the leading anharmonic correction behaves as $r^{-3n}$. Therefore, all terms that decay faster than $r^{-3n}$ can be omitted within the scope of the harmonic approximation. Note also that at the same time all odd moments in Eq. (4) vanish as long as the director field is symmetric about the particle center[2-5]. Thus, a colloidal particle of quadrupole symmetry (Fig. 4a and b) is characterized by a set of three coefficients $b_2$, $b_4$ and $b_6$. In the case of beads with usual planar (Fig. 4a and d) or homeotropic (Fig. 4b and e) boundary conditions, the higher-order terms $b_4$ and $b_6$ are suppressed by the dominant quadrupolar one, $b_2$, so that their influence is substantial only at very short inter-particle distances. However, this analysis does not hold for the particles with conically degenerate surface anchoring. The director in the vicinity of such a bead is a superposition of the two pure quadrupolar configurations separately containing the "Saturn ring" or boojums. For the cases of pure elastic quadrupoles, the former was characterized by[6] $b_2 = 0.4$ while the latter was found to have[2] $b_2 = -0.36$. Thus, the net quadrupole moment of our elastic multipole is small and can be tuned to zero by controlling the conical boundary conditions, so that the higher-order multipoles manifest themselves in a wider range of distances. To verify this scenario, we measured the time dependence of the inter-particle separation between two PSMs (Fig. 3a and Supplementary Fig. 6). The force of interaction between the particles, $\mathbf{F}_{int} = -\nabla U_{int}$, incorporates multipole coefficients $b_2$, $b_4$ and $b_6$ as adjustable parameters, which are found from the experimentally measured $R(t)$. Only a single, unique set of coefficients $b_2$, $b_4$ and $b_6$ provides satisfactory agreement between theoretical and experimental data curves (Fig. 3a and Supplementary Fig. 6), while also explaining the angular dependencies of interactions (Fig. 2).

The analysis of elastic interactions based on multipole expansion has its own limitations. A colloidal particle in the NLC host is surrounded by a small region where the director deviations are large and described by highly nonlinear equations[2]. Elastic interactions in the corresponding range of separations of $R < 2.4r_0$ are, therefore, also characterized by strong nonlinearities. In this region the multipole expansion is not valid and the present theoretical approach does not apply, so we exclude this region from our analysis of experimental data. From fitting experimental data (Fig. 3a and Supplementary Fig. 6), we find that the quadrupole moment $a_2 = b_2 r_0{}^3 = -0.017 r_0{}^3$ is indeed orders of magnitude smaller than that for quadrupolar NLC colloidal spheres with usual tangential or perpendicular surface anchoring. This yields an angular dependence of interactions that is essentially different from that of elastic dipoles and quadrupoles and consists of eight angular sectors of attraction separated by eight sectors of repulsion (Fig. 2a). The interaction energy has two minima in each quadrant (Fig. 2b), with angular positions of minima dependent on the separation $R$. For instance, when $R = 2.4 r_0$ the corresponding minimum-energy angles are approximately 34° and 72° (Fig. 2b). The slight difference between these values and the experimentally measured angles of assemblies $\theta_1 \approx 22°\text{-}26°$ and $\theta_2 \approx 64°\text{-}75°$ is caused by nonlinear distortions concentrated at the region between $r_0$ and about $1.2 r_0$ from the center of every particle, causing complex nonlinear interactions at $R < 2.4 r_0$ that are not captured within the theoretical framework described here.

**Supplementary References**